\documentclass[english]{revtex4}
\usepackage[T1]{fontenc}
\usepackage[latin9]{inputenc}
\usepackage{mathrsfs}
\usepackage{amsmath}
\usepackage{amssymb}

\makeatletter
\@ifundefined{textcolor}{}
{%
 \definecolor{BLACK}{gray}{0}
 \definecolor{WHITE}{gray}{1}
 \definecolor{RED}{rgb}{1,0,0}
 \definecolor{GREEN}{rgb}{0,1,0}
 \definecolor{BLUE}{rgb}{0,0,1}
 \definecolor{CYAN}{cmyk}{1,0,0,0}
 \definecolor{MAGENTA}{cmyk}{0,1,0,0}
 \definecolor{YELLOW}{cmyk}{0,0,1,0}
}

\makeatother

\usepackage{babel}
\begin{document}
\title{Second-Order Moment Quantum Fluctuations and Quantum Equivalence Principle}
\author{M.J.Luo}
\address{Department of Physics, Jiangsu University, Zhenjiang 212013, People's
Republic of China}
\email{mjluo@ujs.edu.cn}

\begin{abstract}
The second-order moment quantum fluctuations or uncertainties are
mass-dependent, and the incompatibility between the quantum uncertainty
principle and the equivalence principle is at the second-order moment
(variation) level, but not the first-order moment (mean) level. To
reconcile the two fundamental principles, we find that the second-order
moment quantum fluctuations are actually distinguished into two parts:
a dynamic part and a geometric part. The dynamic part is indeed mass-dependent
and governed by a non-zero Hamiltonian in a non-general-covariant
inertial frame, and the geometric part is mass-independent and comes
from coarse-graining and/or geometric effects. The dynamic part is
coordinate dependent, it can be canceled away by a coordinate transformation,
and hence it plays no role in general covariant theories whose Hamiltonian
automatically vanishes. However, the geometric part is valid for general
coordinate, and it can not be eliminated by a coordinate transformation.
On the contrary, the geometric part of second-order moment fluctuation
of quantum spacetime leads to coordinate transformation anomaly, which
induces an effective Einstein's gravity theory. The geometric part
is mass-independent and universal, so it is only this part measures
the universal second-order moment quantum fluctuation of the spacetime,
while the dynamic part plays no role in the general covariant description.
The observation generalizes the classical equivalence principle to
the quantum level. And according to the principle, a general covariant
theory with only geometric part quantum fluctuation, i.e. a non-linear
sigma model, is proposed as a theory of a material quantum reference
frame system. The effects of the universal second-order moment quantum
fluctuations in the material quantum reference system and its implications
to an effective gravity theory are also discussed.
\end{abstract}
\maketitle

\section{Introduction}

The classical equivalence principle ensures the symmetry between the
material particles and the material reference frames, so that the
relative motion between them satisfies the principle of relativity.
The principle is the physical foundation of the geometrization of
gravity. Therefore, the current mainstream candidates for geometric
quantum gravity are also based on the assumption that the equivalence
principle is also valid at the quantum level. However, the mass-dependence
of a particle's free-fall given by the standard Schrödinger's quantum
treatment is in serious conflict with the mass-independence of the
free-fall claimed by the equivalence principle. It seems that, at
the quantum level, there is no such symmetry between the particle
and material reference frame because of their general arbitrary masses.
This phenomenon can be seen easily as follows, the quantum uncertainty
principle is the uncertainty between the position $X$ of particles
and their canonical momentum $p$ (which depends on the mass $m$).
Therefore, the evolution amplitude or uncertainty or fluctuation or
broadening of the particle's wave function is generally dependent
on the mass of the particle, e.g. $\langle\delta X^{2}\rangle=\langle X^{2}\rangle-\langle X\rangle^{2}\propto\hbar/m$,
i.e. the larger the mass the smaller the quantum uncertainty. In fact,
the mass dependence of quantum fluctuation is rather general and inherent
in the usual quantum physics, not only appearing in the non-relativistic
quantum mechanics but also in relativistic quantum field theories.
Without loss of generality, we take the relativistic Klein-Gordon
field $\phi$ with mass $m$ as an example, in this case, the second-order
moment fluctuation of the single-particle field $\phi(x)$ at a spacetime
point $x$ is given by the two-point Feynmann propagator $K\left[\phi^{\prime}(x^{\prime}),\phi(x)\right]$
at the identical point limit, i.e. $\lim_{x^{\prime}\rightarrow x}K\left[\phi^{\prime}(x^{\prime}),\phi(x)\right]=\lim_{x^{\prime}\rightarrow x}\left[\langle\phi^{\prime}(x^{\prime})\phi(x)\rangle-\langle\phi^{\prime}(x^{\prime})\rangle\langle\phi(x)\rangle\right]=\lim_{x^{\prime}\rightarrow x}\int\frac{d^{4}p}{(2\pi)^{4}}\frac{1}{p^{2}-m^{2}+i\epsilon}e^{-ip\cdot(x^{\prime}-x)}$,
in which the mass $m$ explicitly shows up in the integral. In this
sense, although material reference frames with different masses have
mass-independent geodesics for the classical measurement of spacetime,
the uncertainty or fluctuations in material reference frames with
different masses cannot be used to measure the universal spacetime
uncertainty or fluctuations. The quantum uncertainty principle is
in fact a statement about the second-order moment (variance), $\langle\delta X^{2}\rangle$
or $\langle\delta p^{2}\rangle$ or $\langle\phi^{2}(x)\rangle-\langle\phi(x)\rangle^{2}$,
around the first-order moment (mean), $\langle X\rangle$ or $\langle p\rangle$
or $\langle\phi(x)\rangle$. More precisely, the contradiction between
the equivalence principle and the uncertainty principle of quantum
mechanics mainly arises in the second-order moment \citep{1997Testing},
while the Ehrenfest theorem ensures that the first-order moment is
compatible with the Newtonian mechanics, and hence supports the classical
equivalence principle. The underlying reason for this is that the
Schrödinger equation views the free-fall of waves as a dynamic evolution
driven by a (mass-dependent) Hamiltonian energy on a fixed ``spacetime
container.'' In contrast, in general relativity and the equivalence
principle, the free-fall of waves can be equivalently seen as the
waves themselves just sit still while the universal ``spacetime container''
that hosts them undergoes universal changes. These two perspectives
on the relative free-fall (of the waves and the spacetime container)
exhibit asymmetry or inequivalence when considered within the frameworks
of standard quantum theory and general relativity. Firstly, in the
quantum theoretical framework, waves possess second-order moment quantum
uncertainties, whereas the classical ``spacetime container'' lacks
a corresponding symmetric second-order moment quantum uncertainty.
Secondly, even if we consider the second-order moment quantum uncertainty
or fluctuation of the material quantum reference frames in the standard
sense of quantum theory, these uncertainties are controlled by mass-dependent
Hamiltonian energies. Therefore, material reference systems with different
masses appear to have different second-order moment quantum uncertainties,
so it is not possible to measure a universal second-order moment quantum
fluctuation of spacetime by using the material quantum reference frame. 

From our perspective, if the equivalence principle is violated at
the quantum level, meaning that rulers or clocks of different masses,
as material reference systems, can no longer measure a universal fluctuating
spacetime due to the absence of a universal second-order moment uncertainty,
this would be a catastrophe for physics which is fundamentally based
on the spacetime measurement. We are faced with two options: either
to start anew and construct a non-universal spacetime theory that
depends on the mass of the reference frame (in which case gravity
would revert to being an ordinary ``force'' as the Newtonian gravity
before the General Relativity), or to reconsider a more general quantum
theory that deals with material reference systems in a way that allows
for a universal measurement of spacetime (in which case gravity remains
the universal geometric property of spacetime, rather than simply
a ``force''). In this paper, we suggest the latter approach. The
mass-independence of material quantum reference systems in a more
general and fundamental quantum theory does not arise from the Schrödinger
dynamics with a non-zero Hamiltonian, because essentially, any quantum
system with a non-zero Hamiltonian-driven evolution implicitly assumes
a preferred inertial frame of reference, which is a concept the General
Relativity aims to discard from the outset. Instead, in this paper
we suggest that it can arise from the coarse-graining process of material
quantum reference frame (a Ricci flow effect of spacetime). The mass-independence
of the material Quantum Reference Frames (QRF) is one of the crucial
features of the quantum equivalence principle (QEP), because when
these material frame fields possess certain universal quantum properties
that are independent of specific matter, such as the ubiquitous second-order
fluctuation broadening and even higher-order fluctuations, the quantum
fluctuations of these frame fields can be abstracted and reinterpreted
as quantum fluctuations inherent to the spacetime geometry itself,
rather than being attributed solely to the quantum fluctuations of
the material quantum fields serving as the frames. Then a QEP serves
as the foundation for constructing a universal physical framework
to measure spacetime through the material QRF fields.

Let us list some experimental and observational evidences of the equivalence
principle for some quantum objects, although they are not-so-complete.
(i) The interferometer of matter waves is used to measure the Eötvös
factor of cold atoms of Rb and K with different masses, and the result
is almost zero ($O(10^{-7})$) \citep{2020Quantum,2022Testing}, which
indicates that atoms with different masses still universally free-fall
with high accuracy at the quantum level. (ii) Moreover, experiments
on the free-fall of quantum antiparticles have also shown almost the
same free-fall as that of ordinary particles \citep{ALPHA:2023dec}.
(iii) In addition to the explanation of quantum phase shift caused
by the gravitational potential added to the Hamiltonian, the phase
shift in the neutron interference experiment in the gravitational
field (Colella-Overhauser-Werner (COW) experiment) can also be consistently
explained by the universal time dilation of neutrons in different
gravitational potentials. Therefore, the COW experiment is also considered
as an indirect quantum test of the equivalence principle \citep{1974Experimental,1975Observation}.
(iv) Although the energies of spectral lines observed in the universe
can vary greatly, at relatively large cosmic distances, when the proper
motion of the light source becomes relatively small compared to the
cosmic expansion comoving, these spectral lines of different energies
at the same distance from the earth that are comoving with the cosmic
expansion exhibit almost the same universal redshift (i.e., spectral
lines with different energies receding with the expanding space as
they universal ``free-fall''). This can be regarded as a certain
degree of verification of the QEP. In addition, the universality of
gravitational redshift (free-fall of photon) has also been verified
to a very accurate level on the earth experiment \citep{1979GReGr..10..181V}.
(v) Although we do not yet have direct measurements of the universal
broadening of the spectral lines, we have indeed observed a universal
deceleration factor $q_{0}\approx-0.64$ measured at the quadratic
order of redshift in the distance-redshift relation of spectral lines.
This factor, independent with the energies of the spectral lines,
is almost uniform and isotropic in the cosmic background, which is
equivalent to a universal second-order moment broadening of spectral
line redshifts and the modification of the distance-redshift relation
at the quadratic order (Appendix B) \citep{Luo2015Dark,Luo:2019iby,Luo:2021zpi}.
So, to a certain extent, it is also an indirect quantum test of the
equivalence principle. (vi) Similar to the thought experiment of the
free-fall in the Leaning Tower of Pisa, consider a heavy ball and
a light ball connected by a rope. If the heavy ball and the light
ball have different falling speeds, there will be a mutual dragging
force between them (this dragging force would slow down the heavy
ball and speed up the light ball, but this contradicts the fact that
when the two balls are combined, their center of mass becomes heavier,
and the overall falling speed should be faster than that of a single
ball). The quantum version of this thought experiment can be correspondingly
considered as a hydrogen atom composed of a relatively heavy proton
and a relatively light electron, which accelerates and receding with
the cosmic background expansion. If the heavy proton and the light
electron have different ``falling'' (receding) speeds in the cosmic
expansion background, then in addition to the Coulomb force between
the proton and the electron, there will be an additional contribution
of a mutual dragging force. This would result in a slight shift in
the measured comoving recession hydrogen atom spectrum compared to
the spectrum in a laboratory inertial frame, or equivalently, a deviation
in the measured fine-structure constant from the laboratory frame
measurement. Therefore, measuring the spectral redshift or fine-structure
constant $\alpha$ of recession hydrogen atoms in the cosmic background
can be used to indirectly test the validity of the QEP. Within the
error range, the current measurements of the fine-structure constant
in the cosmic expansion background show very small ($\Delta\alpha/\alpha\sim O(10^{-5})$)
deviations from the laboratory frame measurements \citep{Wilczynska:2020rxx}.
This can also be regarded, to a certain extent, as a quantum test
of the equivalence principle. Similar idea that the violation of the
equivalence principle may cause corrections in the hyperfine structure
(such as the Lamb shift and anomalous magnetic moment) at the atomic
level, several constraints have been set to the equivalence principle
\citep{1996MPLA...11.1757A,1996PhRvD..54.5954A,1996PhRvD..54.7097A}. 

Past works on the relation between the equivalence principle and the
quantum theory cover a broad range, from the quantum treatment of
free-fall to the discussions of the validity of the equivalence principle
at the quantum level. Because of the vast amount of existing literature,
We can only make a not-so-complete review as follows. The standard
non-relativistic Schrödinger equation treatments of a free-fall wave
function are e.g. given in \citep{Eliezer1977The,1979AmJPh..47..264B,Wadati2000The,Vandegrift2000Accelerating,2012Weak,Nauenberg2016Einstein,2022Free}.
In these papers the gravitation is treated by adding a linear Newtonian
potential (uniform acceleration) to the free Hamiltonian as an ordinary
force, and the Airy function in space coordinate as the solution is
derived. The basic observation from the solution is that the universal
free-fall or the classical weak equivalence principle seems violated.
Further, the observations that the wavefunction and propagator are
explicitly mass-dependent seem to be the direct evidence of the violation
of the weak equivalence principle at the quantum level \citep{Greenberger1970Theory,1983The,2006Quantum,2017Does,Ali:2006ub}.
More detailed studies of the violation of the weak equivalence principle
by Viola and Onofrio \citep{1997Testing} shows that it is the second-order
moment fluctuations around the mean value carry the mass-dependent,
thus it is at the second-order moment signals the non-universality
of the quantum free-fall, but the first-order moment quantum mean
geodesic motion are compatible with the classical equivalence principle
\citep{2004CQGra..21.2761D}. Many works accepted the break down of
the quantum version of the universal free-fall and the equivalence
principle, but some attempts have been proposed to reconcile the mass-dependence
and the equivalence principle at the quantum level. Some of the papers
\citep{Eliezer1977The,Kucha1980Gravitation,Greenberger1998Some,Herdegen:2001tt,2022Free}
try to save the equivalence principle in such a way that the linear
Newtonian potential is argued to be eliminated by a coordinate transformation
from the Newtonian inertial frame to the free-falling frame, but with
an extra local $U(1)$ phase factor in the wavefunction. The phase
factor is also mass-dependent, so the linear Newtonian potential still
can not be universally eliminated, the non-universality of this elimination
effect is just hidden in the extra phase factor. In addition, there
is no evidence for the validity of the Schrödinger equation in the
accelerated falling frame, and there is certainly also no evidence,
at this level, for a unification of the gravitation (the coordinate
transformation of the wavefunction) and the electromagnetism (the
extra local $U(1)$ phase factor of the wavefunction). Greenberger
\citep{Greenberger1970Theory,Greenberger1998Some} notices that the
mass-dependence is rooted in the quantum commutation rule in which
the canonical momentum $p$ is mass-dependent, so he suggests that
to implement the mass-independence and equivalence principle at the
quantum level one should construct an alternate quantum commutation
rule through replacing the canonical momentum $p$ by a velocity $v$,
the price to pay is the need to introduce a fundamental length scale.
Zych and Bruckner \citep{2017Quantum} suggest constraints on the
properties of the energies operator and Hamiltonian of quantum particles
to implement the quantum equivalence principle. Anastopoulos and Hu
\citep{Anastopoulos_2018} suggest that it is the probability distribution
(modulo a mass-dependent phase factor) of free-fall particles is mass-independent
rather than the complex wavefunction or propagating amplitude. Emelyanov
\citep{2022On} argues that the universality of free-fall and the
wavefunction spreading or second-order moment are mutually exclusive
phenomena, and suggests the universality of free-fall is not so fundamental
in the equivalence principle than the wavefunction spreading. Okon
and Callender \citep{Okon:2010kn} emphasized that the statement of
the equivalence principle comes in different varieties, regardless
of whether the free-fall is universal or not, they argue that it does
not preclude the Einstein's equivalence principle, an equivalence
principle should be formulated and valid in a more wider level that
not clash with quantum mechanics. Giacomini, Castro-Ruiz, and Brukner,
etc. \citep{Giacomini:2017zju,2020arXiv201213754G,2022Quantum} suggest
Einstein's equivalence principle could hold for a wider class of reference
frames that are associated to quantum system in a superposition of
spacetime. The author \citep{Luo:2023eqf} finds that when a plane
wavefunction is under a pure coordinate transformation (without an
artifact $U(1)$ phase shift) from the initial frame to a relative
accelerated frame, the acceleration locally contributes an additional
broadening to its spectrum, by using the local Fourier transformation
method (the Gabor transformation), and the extra spectrum broadening
is only given by the local short-time acceleration without any mass-dependence.
In this paper, different from the ones described above, we will develop
a new point of view to the mass-dependence paradox or universality
paradox in the second-order moment quantum fluctuation, so that one
can reconcile the quantum uncertainty principle with the equivalence
principle naturally.

The paper is organized as follows. In section II, we discuss how to
reconcile the apparent confliction of the mass-dependence of the second-order
moment quantum fluctuations with the equivalence principle by noticing
that the fluctuations are in fact distinguished into a dynamic part
and a geometric part. The generalizations of the equivalence principle
to the quantum level are discussed in section III. In section IV,
according to the quantum equivalence principle, we discuss that the
probe of the universal second-order moment quantum fluctuations of
the spacetime must be based on the general covariant theory with only
geometric part fluctuations, which is the idea of material quantum
reference frame system. We conclude the paper in section V. Some supplementaries
on the spacetime Ricci flow due to the universal second-order moment
quantum fluctuation of the quantum reference frame system, the associated
effective gravity theory and the modification to the distance-redshift
relation of the universe are briefly discussed in the Appendix A and
B. 

\section{Reconcile the Mass-dependence of the Quantum Uncertainty with the
Equivalence Principle}

We first concentrate on the problem how to reconcile the textbook
quantum uncertainty principle and the equivalence principle. That
is, why the textbook Schrödinger equation gives rise to a mass-dependent
quantum uncertainty? Does it really clash with the equivalence principle?
And is it possible to achieve a mass-independent quantum uncertainty
at the quantum level that a QEP demands? How these two quantum uncertainties
are both true? What is their relation? To answer the questions, the
key is to notice that the mass-dependence of the quantum uncertainty,
e.g. $\langle\delta X^{2}\rangle\propto\hbar/m$ in non-relativistic
quantum mechanics (or its relativistic single-particle field version
$\langle\phi^{2}(x)\rangle-\langle\phi(x)\rangle^{2}=\int\frac{d^{4}p}{(2\pi)^{4}}\frac{1}{p^{2}-m^{2}+i\epsilon}$,
we also take the Klein-Gordon field as an example in the following
discussions without loss of generality) is given by Schrödinger equation
(or the relativistic quantum field version is by a Schrödinger functional
equation), which is Hamiltonian-driven and hence is only valid in
a non-general-covariant inertial frame (the non-relativistic quantum
mechanics only applies to Galileo's inertial frame and relativistic
quantum field theory only applies to Lorentz's inertial frame, both
have non-trivial Hamiltonian), so it is inertial frame dependent and
non-universal, we call it the ``dynamic part'' of quantum uncertainty. 

The dynamic part of quantum uncertainty is governed by its Hamiltonian.
The classical geodesic trajectory of a test particle is determined
by the variation of the Hamiltonian $\delta H=0$. Since kinetic energy
$T(m_{i})$ is proportional to its inertial mass $m_{i}$, and gravitational
potential energy $V(m_{g})$ is proportional to its gravitational
mass $m_{g}$, and given the condition that $m_{i}=m_{g}=m$, the
classical trajectory $\delta H(m)=\delta T(m)+\delta V(m)=0$ is thus
proportional to mass $m$. Consequently, the mass $m$ can be dropped
out from the classical equation of motion $\delta H(m)=0$, leading
to the mass-independence of geodesic orbits at the classical level.
However, at the quantum level, the variation $\delta H(m)$ now is
the quantum fluctuations, which are generally non-zero in an inertial
frame, rendering $\delta H(m)\neq0$. So, even if the inertial mass
equals the gravitational mass, gravitational effects at the quantum
level appear to be generally mass-dependent. To achieve the mass-independence
in the dynamic part requires a much stronger condition than $m_{i}=m_{g}$. 

Because the quantum geodesic amplitude $K(\hat{X},X,s)=\langle\hat{X}|e^{-isH(X,p)}|X\rangle$
(or its relativistic field version $K\left[\hat{\phi}(\hat{x}),\phi(x),s\right]=\langle\hat{\phi}(\hat{x})|e^{-isH[\phi,\dot{\phi}]}|\phi(x)\rangle$)
encodes almost all quantum information of a particle's (or relativistic
field's) propagation (``geodesic'' means that there is no other
interactions on the particle, except gravitation, e.g. free-fall),
$K(\hat{x},x,s)$ (or $K\left[\hat{\phi}(\hat{x}),\phi(x),s\right]$)
must play the role of the generalization of the classical geodesic
trajectory $l(\hat{X},X,s)$ of a particle which is mass-independent
(for $l(\hat{X},X,s)$ is the solution of the geodesic equation $\delta H(X,p)=0$).
However, in inertial frame, the integration $\int_{0}^{\infty}dsK(\hat{X},X,s)$
(or $\int_{0}^{\infty}dsK\left[\hat{\phi}(\hat{x}),\phi(x),s\right]$)
gives rise to the Feynmann propagator $K(\hat{X},X)$ (or $K\left[\hat{\phi}(\hat{x}),\phi(x)\right]$)
which is mass-dependent, so is $K(\hat{X},X,s)$ (or $K\left[\hat{\phi}(\hat{x}),\phi(x),s\right]$).
Note that $K(\hat{X},X,s)$ (or $K\left[\hat{\phi}(\hat{x}),\phi(x),s\right]$)
has summed over all classical trajectories (or all relativistic field
configurations) connecting the spacetime points from $X$ to $\hat{X}$
(or configurations from $\phi(x)$ to $\hat{\phi}(\hat{x})$) with
an external parameter $s$ (often called the ``proper time''), therefore,
to make the geodesic amplitude $K(\hat{X},X,s)$ (or $K\left[\hat{\phi}(\hat{x}),\phi(x),s\right]$)
independent of mass, just like the classical trajectory equation $\delta H(X,p)=0$
(or relativistic field equation $\delta H(\phi,\dot{\phi})=0$), which
can eliminate the mass $m$ inside, the required condition is that
the Hamiltonian on all different trajectories (or all different relativistic
field configurations) should have $H(X,p)=0$ (or relativistic field
version $H(\phi,\dot{\phi})=0$). We will show that the condition
is not only achievable, but also natural. It is called Hamiltonian
constraint, which appears automatically in any general covariant theory.

We see that the mass independence of $K(\hat{X},X,s)$ (or $K\left[\hat{\phi}(\hat{x}),\phi(x),s\right]$)
occurs in such a way that the proper time parameter $s$ of a classical,
absolute, external, inertial frame of reference is unphysical, because
we work on an inherent 4-spacetime $X\,(\textrm{or}\,x)\in M^{4}$,
and there is no external $M^{5}=M^{4}\times s$ and global parameter
$s$. At the same time, there is no concept of classical geodesic
trajectory (or classical relativistic field configuration) at the
quantum level, this proper time parameter $s$ can not be exactly
achieved. Physically, we also know that if $s$ is interpreted as
some kind of external proper time, due to the influence of the limited
speed of light on the ``comparison of proper time clocks in different
local patches'', since the rate of the proper time clocks on different
local coordinate patches are different, and there is no globally unified
proper time parameter $s$ between different local coordinate patches.
The Hamiltonian constraint $H(X\in M^{4},p)=0$ (or $H(\phi,\dot{\phi})=0$)
is actually an inevitable result of the absence of external absolute
time parameters in the general covariant coordinate system at the
quantum level, because in essence, any quantum system with non-zero
Hamiltonian-driven evolution implies a preferred inertial frame or
an external larger global space. In this sense, without the external
parameter $s$ and using the Hamiltonian constraint, we have
\begin{equation}
K(\hat{X},X)=\int_{0}^{\infty}ds\langle\hat{X}|e^{-isH(X,p)}|X\rangle\overset{H=0}{=}\left(\int_{0}^{\infty}ds\right)\langle\hat{X}|X\rangle\propto\int_{0}^{|p|=k}\frac{d^{4}p}{(2\pi)^{4}}e^{-ip\cdot(\hat{X}-X)}\overset{k\rightarrow\infty}{=}\frac{dV(\hat{X})}{dV(X)}\delta^{(4)}(\hat{X}-X),\label{eq:kernel}
\end{equation}
in which $k\in(0,\infty)$ is the cutoff of the Fourier modes of the
amplitude, and $\frac{dV(\hat{X})}{dV(X)}$ is the Jacobian that measures
the local volume change in different coordinate patches. Its relativistic
single-particle (S.P.) (i.e. $\langle\phi(x)|p\rangle=e^{-ip\cdot x}$
having a well-defined momentum) field version of the Feynmann propagator
gives a similar result
\begin{align}
K\left[\hat{\phi}(\hat{x}),\phi(x)\right] & =\int_{0}^{\infty}ds\langle\hat{\phi}(\hat{x})|e^{-isH[\phi,\dot{\phi}]}|\phi(x)\rangle\overset{H=0}{=}\left(\int_{0}^{\infty}ds\right)\langle\hat{\phi}(\hat{x})|\phi(x)\rangle\nonumber \\
 & \overset{S.P.}{\propto}\int_{0}^{|p|=k}\frac{d^{4}p}{(2\pi)^{4}}e^{-ip\cdot(\hat{x}-x)}\overset{k\rightarrow\infty}{=}\frac{dV(\hat{x})}{dV(x)}\delta^{(4)}\left(\hat{x}-x\right).\label{eq:kernel field}
\end{align}
In a non-single-particle case, $K\left[\hat{\phi}(\hat{x}),\phi(x)\right]$
is proportional to the overlap between two states of the field configurations
$\langle\hat{\phi}(\hat{x})|\phi(x)\rangle$ (as a smearing function
formally replacing the Dirac delta) which is also mass-independent.
So we proved that the Hamiltonian constraint is a necessary condition
for the mass-independence of the quantum geodesic amplitude for general
coordinate system.

It is also worth noting that the broadening of eq.(\ref{eq:kernel})
(or eq.(\ref{eq:kernel field})) is not the ``dynamic part'', it
is not Hamiltonian-driven, rather than arises from the coarse-graining
effect of the cutoff scale $k$. When the finite cutoff $k$ is applied,
this broadening is independent of mass and only related to the cutoff.
Since in practical experiment only a finite resolution is available,
the width or fuzziness of $K(\hat{X},X)$ (or $K\left[\hat{\phi}(\hat{x}),\phi(x)\right]$)
depends on the resolution we observe it. The short distance behavior
of the modes beyond the cutoff are coarse-grained. The width of $K(\hat{X},X)$
(or $K\left[\hat{\phi}(\hat{x}),\phi(x)\right]$) changes with the
cutoff, which is just the process of the renormalization flow, we
will make the idea more precisely latter. In the limit as the cutoff
approaches infinity, all Fourier modes are detected and hence we have
infinite resolution, it becomes a delta function without broadening.
We can see that eq.(\ref{eq:kernel}) (or eq.(\ref{eq:kernel field}))
does not depend on the particle mass in a general coordinate system,
nor does it change with the (nonexistent hypothetical) evolution path
parameter $s$, but only depends on the general background spacetime
$X\,(\textrm{or}\,x)\in M^{4}$ through the Jacobian. In essential,
using the kernel (of the test particles) to probe the background spacetime
is just the mathematical problem \citep{1966Can} ``hear the shape
of a drum'': by detecting the sound wave modes on the drum to probes
the geometry of the drum. So, we call this extra coarse-graining-driven
broadening or quantum uncertainty, the ``geometric part'' (compared
with the previous ``dynamic part''), which is mass and coordinate
independent and universal.

As an example, we consider the Non-Linear Sigma Model (NLSM) eq.(\ref{eq:NLSM})
on a $d=4-\epsilon$ base spacetime, which is used in later section
as a model of material QRF. For the reason that it is general covariant
or background independent of the base spacetime, so it is a Hamiltonian
constraint system. As given by the equivalence principle, it is well
known that its classical equation of motion of NLSM is a geodesic
equation which is mass-independent. As demanded by the QEP, the (inverse
of) quantum uncertainty or broadening of the target spacetime of NLSM
eq.(\ref{eq:NLSM}) given by the Ricci flow eq.(\ref{eq:Ricci flow})
(a coarse-graining effect at Gaussian approximation) is cutoff-dependent
but mass-independent \citep{Luo:2019iby,Luo:2021zpi}
\begin{equation}
\langle\delta X^{\mu}\delta X^{\nu}\rangle=\frac{1}{\lambda}g^{\mu\nu}\int_{0}^{|p|=k}\frac{d^{d}p}{(2\pi)^{d}}\frac{1}{p^{2}}=\frac{1}{\lambda}g^{\mu\nu}\frac{k^{d-2}}{(4\pi)^{d/2}\Gamma(d/2+1)}\overset{d\rightarrow4-\epsilon}{=}\frac{k^{2}}{32\pi^{2}\lambda}g^{\mu\nu}\label{eq:2nd moment}
\end{equation}
in which $\lambda$ is a universal coupling constant of the NLSM,
$g_{\mu\nu}$ the geometric metric of the target spacetime. A key
feature of NLSM worth stressing is that, gravity is not introduced
by adding a Newtonian gravitational potential to its Hamiltonian (as
the Schrödinger equation do), but is covariantly introduced through
the geometric metric. This is suggested as the fundamental reason
why the textbook Schrödinger equation gives rise to a mass-dependent
quantum uncertainty, because in this sense, the Schrödinger equation
with the Newtonian potential for a particle is not precisely geodesic
free-fall. Such particles are simultaneously endowed with two parts
of the quantum uncertainties: the non-universal ``dynamic part''
quantum uncertainty due to their own proper motion relative to the
inertial frame where the non-zero Hamiltonian defined, and the universal
``geometric part'' that encodes in the coarse-grained free-fall
particle or symmetrically in the material reference frame without
relative to any inertial frame, which is although not dominant in
this case. Instead, a precise free-fall should be described by a generally
coordinate invariant Hamiltonian constraint system (as the NLSM),
in which there is only the universal ``geometric part'' quantum
uncertainty, without the ``dynamic part'' of the particle's own
proper motion relative to any inertial frames.

To sum up, the quantum second-order moment fluctuations/quantum uncertainty
of material particles (also for the material quantum reference systems)
generally stem from two components. If there is a non-zero Hamiltonian
in a special inertial frame, it gives rise to the non-universal ``dynamic
part'' by Schrödinger equation (therefore, this broadening cannot
be interpreted as a universal spacetime broadening, similar to the
fact that the various redshifts caused by the peculiar proper motions,
relative to the co-expansion frame, of celestial bodies cannot be
explained as the universal cosmic background expansion, but solely
their individual proper motions). While in general coordinate invariant
theories, due to a Hamiltonian constraint, there is no such contribution.
However, there is always a non-dynamical, coarse-grained and purely
``geometric part'' (as the Ricci flow of NLSM) contribution, which
can be extracted and reinterpreted as a universal spacetime fluctuation
or broadening (similar to the fact that the cosmological redshift
caused by the common cosmic recessional motion of celestial bodies
can be explained as the universal cosmic spatial expansion, also called
the Hubble flow, when their individual proper motions are relatively
small at cosmic distances). In typical laboratory settings, the former
broadening is dominant (for instance, this broadening can be explained
in terms of the lifetimes of various particles or states), while the
latter is relatively insignificant, allowing spacetime to be considered
classical. When the latter uncertainty becomes significant (e.g. at
cosmological distances, where the universal second-order moment broadening
of the comoving spectral lines leads to second-order corrections in
the distance-redshift relation and can be explained as the universal
cosmic expansion acceleration), the quantum nature of spacetime cannot
be ignored. 

\section{Quantum Equivalence Principle}

Based on the results in previous section, we find that the second-order
moment quantum fluctuation or quantum uncertainty does not really
clash with the equivalence principle, if one notices that it is the
universal ``geometric part'' of the quantum uncertainty measures
the quantum fluctuation of spacetime. But the non-universal ``dynamic
part'' $\langle\delta X^{2}\rangle\propto\hbar/m$ (or its relativistic
field version $\langle\phi^{2}(x)\rangle-\langle\phi(x)\rangle^{2}$)
is defined in a non-general-covariant inertial frame and hence can
always be eliminated by a coordinate transformation to nullify the
Hamiltonian in the inertial frame. Consequently, the notorious mass-dependence
of the second-order moment and quantum geodesic amplitude are also
eliminated. These facts give us physical foundations to formulate
a generalized equivalence principle at the quantum level. According
to various formulations of the classical equivalence principle, here
we give some generalizations. 

\subsection{Quantum Universal free-fall (Weak Equivalence Principle)}

\paragraph{Classical: }

All test bodies free-fall in a gravitational field with the same geodesic
trajectory (position, velocity and acceleration) regardless of their
masses or internal composition.

\paragraph{Quantum: }

All test scalar particles (without internal degrees of freedom) free-fall
in a gravitational field with the same the quantum geodesic amplitude
(not only the first-order moment, but also the geometric part of second-order
moment, possibly even higher order moments) regardless of their masses. 

First, we suggest to replace the general bodies by the scalar particles
without internal structure. It is more precisely known that spinning
objects/particles or composite particles do not exactly follow classical
geodesics \citep{2015Spinning} and hence do not exactly quantum free-fall.
Scalar structure-less particles (e.g. NLSM) are sufficient to serve
as an intermediate equivalent for comparing different motions, more
precisely, they are enough to be quantum frame fields to probe or
measure the (quantum) Riemannian nature of spacetime geometry (NLSM's
target spacetime) without torsion. Second, we also suggest to replace
the classical geodesic trajectory $l(\hat{X},X,s)$ by a quantum geodesic
amplitude $K(\hat{X},X)$ (or $K\left[\hat{\phi}(\hat{x}),\phi(x)\right]$)
, which has summed over all possible trajectories (or field configurations)
and without the external parameter $s$. Analogous to that one can
extract classical variables such as position, velocity and acceleration
from a trajectory, one can instead extract quantum variables such
as the first-order moment (mean value of position or velocity or relativistic
field value) and second-order moment from the amplitude, and in which
the acceleration of a particle just corresponds to the geometric part
of the second-order moment (additional broadening or smearing of its
spectrum) \citep{Luo:2023eqf}. It is the findings in section II ensures
that the Weak Equivalence Principle still holds at the quantum level
for the universal part, i.e. the geometric part, of the second-order
moment quantum fluctuation in general coordinate. 

\subsection{Strong Equivalence Principle}

The Weak Quantum Equivalence Principle focuses on the law of motion
(classical geodesic trajectory or quantum geodesic amplitude), while
the Strong Quantum Equivalence Principle makes a stronger statement
on the general law of nature including the quantum law.

\paragraph{Classical: }

In general spacetime and arbitrary gravitational field, it is possible
to choose a local inertial reference system, with respect to which
all classical laws of nature take the same form as in a frame that
without spacetime curvature, gravity or acceleration.

\paragraph{Quantum: }

In general spacetime and arbitrary gravitational field, it is possible
to choose a local (material) quantum reference system, with respect
to which all law including the quantum laws of nature take the same
form as in a frame without spacetime curvature, gravity, acceleration,
spacetime fluctuation/fuzziness and spacetime thermal effects.

First, we suggest to replace the classical local inertial frame by
a local material quantum reference system. The materialized quantum
reference system is always quantum vibrating and hence generally takes
universal ``geometric part'' quantum fluctuations as ordinary material
particles. In this sense, a classical, external, absolute rest or
inertial frame of reference is unphysical at the quantum level. Due
to this fact, it can help us understand the Hamiltonian-constraint-nature
of a general covariant under-studied system, and why the quantum geodesic
amplitude eq.(\ref{eq:kernel}) (or eq.(\ref{eq:kernel field})) related
to the Hamiltonian constraint is independent of mass, w.r.t. a ``rest''
quantum reference frame (here the quotation mark on the ``rest''
represents that it is just classically at rest relative to the under-studied
system, rather than absolute rest at the quantum level, more precisely,
the first-order moment mean coordinate of the frame is at rest, but
it takes a non-vanishing second-order moment quantum variation around
the mean value, i.e. fluctuating frames are in quantum superposition).
It can also help us understand why the quantum vacuum zero-point fluctuation
energy does not contribute to the cosmological constant \citep{Luo2014The}.
Because, as a dynamic part fluctuation, the zero-point fluctuation
of vacuum vanishes w.r.t. an observer who is standing on a ``rest''
quantum frame. In other words, the zero-point fluctuation of vacuum
can only be observed in an absolute external classical inertial frame,
which has been abandoned in the QEP. The only observed vacuum fluctuation
related to the cosmological constant is the geometric part.

Second, some new acceleration-induced, coarse-graining-induced and
geometric-induced quantum effects can also be introduced or canceled
away by anomaly cancellation or proper quantum coordinate transformation,
just like the classical principle states that the spacetime curvature,
gravity and acceleration can be introduced or canceled away by proper
classical coordinate transformation. Similarly, possible new effects
are also added: (i) quantum spacetime Gaussian fuzziness (spacetime
Ricci flow from the universal second-order moment fluctuation of the
geometric part \citep{Luo:2019iby,Luo:2021zpi,Luo:2022goc,Luo:2022statistics,Luo:2022ywl,2023AnPhy.45869452L},
see Appendix A), (ii) short time acceleration-induced spectrum broadening,
(iii) acceleration-induced Unruh thermal effect, etc. 

The effect (i) is a coarse-graining-induced quantum effect coming
the geometric part fluctuation, it is irreversible because of the
irreversibility of the spacetime Ricci flow (i.e. the H-theorem eq.
(\ref{eq:H-theorem}) of quantum spacetime \citep{Luo:2022statistics}).
The cancellation of the anomaly coming from the effect (i) (known
as the diffeomorphism anomaly cancellation \citep{Luo:2021zpi}) in
a laboratory frame requires a counter term which is related to the
cosmological constant \citep{Luo2014The,Luo:2015pca,Luo:2021zpi}
(see Appendix A). The finite cutoff $k$ broadening of the delta function
eq.(\ref{eq:kernel}) (or eq.(\ref{eq:kernel field})) and the cancellation
of it also belong to this type. The effect (ii), the acceleration-induced
spectrum broadening can also be equivalently understood as another
form of the spacetime fuzziness (effect (i)). For this reason, the
extra redshift broadening modification at the quadratic order of the
distance-redshift relation can also be equivalently interpreted as
the cosmic expanding acceleration (the deceleration parameter $q_{0}$)
(Appendix B) \citep{Luo2015Dark,Luo:2019iby,Luo:2021zpi}. The effect
(ii) and (iii) are both acceleration-induced quantum effect, some
literature argue that the effect (iii), thermal effect, implies the
existence of the absolute inertial frame in quantum mechanics, but
it is shown \citep{Luo:2023eqf} that the effect (ii) and (iii) can
be interpreted as the same type of effects coming from a coordinate
transformation from initial frame to an accelerated frame, the only
difference are effect (iii) is a long-time acceleration with maximum
entropy and (ii) is a short-time acceleration with least information
lost. 

To sum up, at the quantum level, not only the spacetime curvature,
gravity and acceleration, the three things are equivalent, but also
the spacetime curvature, gravity, acceleration, spacetime Gaussian
fuzziness and spacetime thermal effects, these five things are also
equivalent.

\section{Material Quantum Reference System}

In classical physics, the spacetime is an abstract container where
events happen. However, we essentially have never ``really'' measured
the spacetime itself. All our measurements of the spacetime are indirect
measurements by virtue of some material reference systems (such as
light ray or other matter fields as reference systems). The basic
idea of the material quantum reference system which probes the spacetime
geometry is also in analogous to the mathematical problem \citep{1966Can}:
``hear the shape the drum'' by using the universal (first, second,
and even higher order moments) of the modes of the material quantum
frame fields. As discussed in previous sections, we can see that the
non-universal (particle mass-dependent) ``dynamic part'' of quantum
fluctuations arising from a material particle with a non-zero Hamiltonian
do not quantify the intrinsic fluctuations of spacetime itself. According
to the QEP, the quantum dynamical fluctuation or broadening governed
by the system's Hamiltonian can always be transformed away, as the
Hamiltonian of a particle without internal degrees of freedom can
always be transformed into zero in a general covariant description,
constituting a Hamiltonian constraint. To identify a coordinate-independent
and general covariant material quantum reference system that describes
the universal quantum fluctuations of spacetime (irrespective of the
mass of the material quantum reference frame), we introduce a $d=4-\epsilon$
Non-Linear Sigma Model (NLSM) to depict the frame field theory of
the material Quantum Reference Frame \citep{Luo:2015pca,Luo:2019iby,Luo:2021zpi,Luo:2022statistics},
\begin{equation}
S_{X}=\frac{1}{2}\lambda\int d^{d}x\sqrt{|\mathfrak{g}|}\mathfrak{g}_{ab}g_{\mu\nu}\frac{\partial X^{\mu}}{\partial x_{a}}\frac{\partial X^{\nu}}{\partial x_{b}}\label{eq:NLSM}
\end{equation}
The NLSM is a fields theory which maps from a base space $x$ to a
target space $X$. $d=4-\epsilon$ is the dimension of the base space
$x_{a}$ ($a=0,1,...d-1$), which is in general metric $\mathfrak{g}_{ab}$,
and without losing generality it can be interpreted as a flat ($\mathfrak{g}_{ab}=\eta_{ab}$)
laboratory frame coordinate. $X^{\mu}=(X^{0},X^{1},X^{2},X^{3})\in M^{4}$
is the target space in general metric $g_{\mu\nu}$, which is interpreted
as the frame fields living in the laboratory frame and measuring the
general physical spacetime coordinates. $\lambda$ is the coupling
constant taking the value of the critical density eq.(\ref{eq:critial density})
of the universe. 

The NLSM with general $\mathfrak{g}_{ab}$ and $g_{\mu\nu}$ inherently
possesses a zero Hamiltonian, which holds true in any general coordinate
system (without the need for a transformation to the ``rest'' frame
to nullify the Hamiltonian). Its classical equations of motion manifest
as geodesic equations independent of mass, with gravitational effects
introduced geometrically (free-fall independent of particle mass)
rather than by incorporating Newtonian potentials into the Hamiltonian.

When one wants to study a quantum system, for instance, a scalar field
$\phi$, which is considered relative to the material quantum reference
system $X$. The state of the whole system is an entangled state $|\Psi\rangle=\sum_{x}C_{x}|\phi(x)\rangle\otimes|X(x)\rangle$
in the Hilbert space $\mathscr{H}_{\phi}\otimes\mathscr{H}_{X}$,
in which $x$ is the laboratory frame coordinate that the frame fields
and the scalar field both live in. The action of the whole system,
i.e. the scalar field and the material quantum reference frame, is
given by a direct sum of both actions,
\begin{equation}
S[\phi,X]=S_{\phi}+S_{X}=\int d^{d}x\sqrt{|\mathfrak{g}|}\left[\frac{1}{2}\mathfrak{g}_{ab}\frac{\partial\phi(x)}{\partial x_{a}}\frac{\partial\phi(x)}{\partial x_{b}}-V(\phi)+\frac{1}{2}\lambda\mathfrak{g}_{ab}g_{\mu\nu}\frac{\partial X^{\mu}}{\partial x_{a}}\frac{\partial X^{\nu}}{\partial x_{b}}\right]
\end{equation}
where $S_{\phi}=\int d^{4}x\sqrt{|\mathfrak{g}|}\left[\frac{1}{2}\mathfrak{g}_{ab}\frac{\partial\phi}{\partial x_{a}}\frac{\partial\phi}{\partial x_{b}}-V(\phi)\right]$
is the action of $\phi$ w.r.t. the laboratory frame $x$. When the
quantum frame fields can be treated semi-classically, i.e. ignore
their second-order moment quantum fluctuation $\langle\delta X^{2}\rangle$
while only consider their first-order moment mean value $\langle X\rangle$,
the action can be rewritten by a semi-classical approximation or first-order
moment mean fields approximation
\begin{equation}
S[\phi,X]\overset{1st}{\approx}S[\phi(X)]=\int d^{4}X\left\Vert \frac{\partial x}{\partial X}\right\Vert \left[\frac{1}{2}g^{\mu\nu}\frac{\delta\phi(X)}{\delta X^{\mu}}\frac{\delta\phi(X)}{\delta X^{\nu}}-V(\phi)+2\lambda\right]\label{eq:1st action}
\end{equation}
in which $\frac{1}{4}\left\langle \mathfrak{g}_{ab}g_{\mu\nu}\frac{\partial X^{\mu}}{\partial x_{a}}\frac{\partial X^{\nu}}{\partial x_{b}}\right\rangle =\frac{1}{4}\left\langle g_{\mu\nu}g^{\mu\nu}\right\rangle =\frac{D}{4}\equiv1$
is used. The semi-classical action is just the original scalar field
action $S_{\phi}$ up to a constant $2\lambda$, and w.r.t. the general
frame fields $X$ rather than the laboratory coordinates $x$. The
derivative $\frac{\partial}{\partial x_{a}}$ is replaced by the functional
derivative $\frac{\delta}{\delta X^{\mu}}$, the measure $d^{d}x$
is replaced by $d^{4}X$ with a Jacobian $\left\Vert \frac{\partial x}{\partial X}\right\Vert =\sqrt{\left|\det g\right|}$.
For the reason that the Jacobian given by a squared matrix, so the
dimension of $x_{a}$, $d$, at the semi-classical approximation must
be the number of least frame fields $X$, so $d$ must be very close
to $D\equiv4$. That is the reason we choose $d=4-\epsilon$ as the
dimension of the base space of NLSM at the semi-classical level. However,
at the quantum level, $d$ does not necessarily be $4$ because of
dimensional anomaly. And the reason that $d$ can not be exactly $4$
is also for the topological consideration that the homotopic group
for the continuous differentiable mapping $X(x):x\rightarrow X$,
for instance, $\pi_{d}(M^{4})$ is trivial for $d<4$, so that the
mapping $X(x)$ is free from intrinsic topological obstacles and well
defined even at the quantum level. It is also for this reason, the
NLSM is renormalizable in $2\le d<4$, perturbatively ($d=2$) or
non-perturbatively ($2<d<4$).

When the frame fields $X^{\mu}$ are promoted to be quantum fields
with un-ignorable second-order moment quantum fluctuation $\langle\delta X^{2}\rangle$,
we need to go beyond the semi-classical approximation. In this situation,
the QEP ensures that we can probe and measure the quantum properties
of spacetime (second-order moment and possibly even higher order)
through this material quantum reference system, which we leave them
briefly in the Appendix A and some references. 

\section{Conclusion}

In this paper, we develop a new point of view to the mass-dependence
paradox or universality paradox in the second-order moment quantum
fluctuation, so that one can reconcile the quantum uncertainty principle
with the equivalence principle. We find that there are two mutually
independent parts of the second-order moment fluctuations in a quantum
particle/field system. The dynamic part is mass dependent and governed
by a non-zero Hamiltonian in an inertial frame, and the geometric
part is mass independent and comes from coarse-graining and/or geometric
effects. The dynamic part is coordinate dependent, so it can be canceled
away by a coordinate transformation, and hence it plays no role in
general covariant theories whose Hamiltonian vanishes. However, the
geometric part is valid for general coordinate, and hence it can not
be canceled by a coordinate transformation. On the contrary, the geometric
part of second-order moment fluctuation of quantum spacetime leads
to the coordinate transformation anomaly (e.g. the diffeomorphism
anomaly), which induces an effective Einstein's gravity theory (see
Appendix A). The anomaly can only be canceled by counter term, which
is related to a cosmological constant. The geometric part is mass-independent
and universal, so it is only this part measures the universal second-order
moment quantum fluctuation of the spacetime through a material quantum
reference system, while the dynamic part plays no role in a general
covariant description. The observation allows us to generalize the
classical equivalence principle to a quantum version.

This argument explains why a falling particle in an external gravitational
field described by the Schrödinger equation (with non-zero Hamiltonian)
seems mass-dependent and hence seems violate the Equivalence Principle.
A mass-independent quantum treatment of free-fall is proposed to be
described by a general covariant theories automatically satisfying
a Hamiltonian constraint. For example, the Non-Linear Sigma Model
(NLSM), in which the incorporation of gravitation is not by introducing
the Newtonian gravitational potential into its Hamiltonian (its Hamiltonian
trivially vanishes), rather than geometrically and covariantly introducing
in the metric. In this correct quantum treatment of the free-fall,
only the universal geometric part of quantum fluctuation appears. 

This argument and observation remove a notorious stumbling block for
the generalization of the equivalence principle to the quantum level.
The quantum equivalence principle generalized the classical equivalence
from the first-order moment (classical spacetime curve, gravity and
acceleration equivalence) to the second-order moment, possibly even
to higher order. Besides the classical equivalence between spacetime
curvature, gravity and acceleration, at the second-order or quantum
level, several new dimensions of equivalences are also inevitable,
such as the spacetime Gaussian fuzziness (finite entropy) and the
spacetime thermal effect (equilibrium state with maximum entropy).
We also emphasize the importance of the universal second-order moment
quantum fluctuation in the quantum equivalence principle and quantum
gravity. According the quantum equivalence principle suggested in
the paper, a general covariant theory with only the geometric part
quantum fluctuations, i.e. NLSM, can be used to describe a theory
of material quantum reference frame system, which measures the universal
quantum properties of spacetime at least to the second-order moment.
The spacetime Ricci flow coming from the universal second-order moment
quantum fluctuation and the associated effective gravity emerges from
the theory are also briefly given in the Appendix A.
\begin{acknowledgments}
This work was supported in part by the National Science Foundation
of China (NSFC) under Grant No.11205149, and the Scientific Research
Foundation of Jiangsu University for Young Scholars under Grant No.15JDG153.
\end{acknowledgments}

\section*{Appendix A: A Brief Introduction to the Spacetime Ricci Flow from
the Universal Second-Order Moment Quantum Fluctuation}

In this appendix, we briefly overview the Ricci flow of spacetime
associated with the geometric part second-order moment quantum fluctuation
of the material reference frame system, i.e. NLSM, the detailed discussions
can be found in references \citep{Luo2014The,Luo2015Dark,Luo:2015pca,Luo:2019iby,Luo:2021zpi,Luo:2022goc,Luo:2022statistics,Luo:2022ywl,Luo:2023eqf,2023AnPhy.45869452L,2024chinarxiv}. 

The universal second-order moment quantum fluctuation of the spacetime
eq.(\ref{eq:2nd moment}) leads to a coarse-graining or renormalization
process of spacetime at the Gaussian approximation, known as the Ricci
flow,
\begin{equation}
\frac{\partial g_{\mu\nu}}{\partial t}=-2R_{\mu\nu}\label{eq:Ricci flow}
\end{equation}
where $t=-\frac{1}{64\pi^{2}\lambda}k^{2}$ related to the cutoff
$k$ is the Ricci-flow parameter and $R_{\mu\nu}$ is the Ricci curvature
of the spacetime. The Ricci flow equation tells us how the universal
second-order moment quantum fluctuation gradually deforms the metric
of quantum spacetime. Many important problems of quantum gravity can
be derived from the Ricci flow equation and its modified version,
the Ricci-DeTurck flow equation
\begin{equation}
\frac{\partial g_{\mu\nu}}{\partial t}=-2\left(R_{\mu\nu}-\nabla_{\mu}\nabla_{\nu}\log u\right)\label{eq:Ricci-DeTurck}
\end{equation}
which is equivalent to the Ricci flow equation up to a diffeomorphism
to the Ricci curvature. $u(X,\tau)$ is called the manifold density
of the spacetime point $X$, with the normalization condition $\lambda\int dV(X)\,u(X,\tau)=1$,
where $dV(X)=d^{4}X\sqrt{\left|\det g(X)\right|}$ is the volume element.
By using the flow equation (\ref{eq:Ricci-DeTurck}), it satisfies
the conjugate heat equation
\begin{equation}
\frac{\partial u}{\partial\tau}=\left(\varDelta-R\right)u\label{eq:conjugate heat eq}
\end{equation}
where $\tau=t_{*}-t$ is a backward flow parameter, $t_{*}$ is certain
singular scale of the flow, $\varDelta$ is the Laplacian of the spacetime,
$R$ the scalar curvature. The manifold density $u(X,\tau)$ is the
probability density of the frame fields $X_{\mu}$ ensemble, and $K(\hat{X},X,\tau)$
is the kernel of the conjugate heat equation (\ref{eq:conjugate heat eq})
satisfying the initial condition 
\begin{equation}
\lim_{\tau\rightarrow0}K(\hat{X},X,\tau)=u_{0}(X)\delta^{(4)}(\hat{X}-X)\label{eq:initial kernel}
\end{equation}
$K(\hat{X},X,\tau)$ is also the quantum geodesic amplitude of the
frame fields eq.(\ref{eq:kernel}), and it can also be calculated
by the standard path integral method by using the action of NLSM eq.(\ref{eq:NLSM}).
It is worth stressing that the backward flow parameter $\tau$ of
$K(\hat{X},X,\tau)=\langle\hat{X}_{\tau}|e^{\tau(\varDelta-R)}|X\rangle$
in general coordinates, is not the ``proper time'' $s$ of $K(\hat{X},X,s)=\langle\hat{X}_{s}|e^{-isH}|X\rangle$
in inertial frame mentioned in the section II. $\tau$ evolution is
a coarse-graining-driven (here is $\varDelta-R$ driven or volume-change
driven) process that produces the geometric part broadening to the
kernel, but $s$ evolution is Hamiltonian-driven process that produces
the dynamic part, which does not exist in the general covariant quantum
reference frame theory.

Compared eq.(\ref{eq:initial kernel}) and eq.(\ref{eq:kernel}) (or
eq.(\ref{eq:kernel field})), or by considering its normalization
condition, the manifold density can also be interpreted as the volume
ratio or Jacobian. The conjugate heat equation is a flow equation
evolves and broadens the initial delta function eq.(\ref{eq:initial kernel})
or eq.(\ref{eq:kernel}) to be a heat-kernel-like solution with non-trivial
broadening. 

Because the QEP ensures a symmetric role between the object and the
material quantum frame field, in this sense, the kernel $K(\hat{X},X,\tau)$
not only describes an object's geodesic amplitude between spacetime
point $X$ and $\hat{X}$, but also a transition amplitude between
the different frame configurations. As a consequence, a general quantum
coordinate transformation, for a field $\phi$, from a coordinate
system $X$ to a fuzzy one $\hat{X}_{\tau}$ at the coarse-graining
scale given by $\tau$, is defined as an integral transformation by
using the kernel $K(\hat{X},X,\tau)$, i.e. 
\begin{equation}
\phi(\hat{X}_{\tau})=\int dV(X)\,K(\hat{X},X,\tau)\,\phi(X)\label{eq:integral transformation}
\end{equation}
Such a quantum coordinate transformation is clearly independent to
the mass of the frame fields $X$ and $\hat{X}$, for the reason that
the kernel only carries the geometric part fluctuations as shown before.
In the sense that the kernel has summed over all classical trajectories,
the quantum coordinate transformation is a quantum superposition of
all classical coordinate transformations.

In a special case when $X$ and $\hat{X}$ are at identical scale,
i.e. $\tau\rightarrow0$, the delta function of eq.(\ref{eq:initial kernel})
or eq.(\ref{eq:kernel}) (or eq.(\ref{eq:kernel field})) makes the
quantum coordinate transformation eq.(\ref{eq:integral transformation})
recover the classical coordinate transformation, i.e. it is just a
trivial coordinate replacement $X\rightarrow\hat{X}$, with a possible
local volume change between the coordinates given by the Jacobian
$\frac{dV(\hat{X})}{dV(X)}=u_{0}$. However, different from the classical
coordinate transformation and other proposed quantum coordinate transformation
\citep{Giacomini:2017zju,2020arXiv201213754G,2022Quantum}, eq.(\ref{eq:integral transformation})
is irreversible and non-unitary under the flow parameter $\tau$.
In certain sense, the quantum coordinate transformation eq.(\ref{eq:integral transformation})
is the mean value coordinate (first-order) classical transformation
in addition to the coordinate blurring (broadening/variance) around
the mean value (second-order), possibly even higher order (skewness
and kurtosis...) corrections.

The irreversibility of the Ricci flow and the related irreversible
fuzziness of spacetime under a quantum coordinate transformation is
demonstrated by a monotonicity of a Shannon entropy in terms of the
manifold density $u(X)$
\begin{equation}
N(M^{4},\tau)=-\int_{M^{4}}dV(X)\,u(X,\tau)\log u(X,\tau)\label{eq:shannon entropy}
\end{equation}
There is an analogous monotonic H-theorem \citep{Luo:2022statistics}
(in analogous to the H-theorem for dilute gas of Boltzmann)
\begin{equation}
\frac{d\tilde{N}}{d\tau}\le0\label{eq:H-theorem}
\end{equation}
for the relative Shannon entropy $\tilde{N}(M^{4},\tau)=N(M^{4},\tau)-N(M_{*}^{4})$
where $N(M_{*}^{4})$ is an extreme value of Shannon entropy given
by an equilibrium manifold density $u_{*}$ at the singular scale
$t_{*}$. Or $\tilde{N}(M^{4},\tau)$ is equivalently the Shannon
entropy eq.(\ref{eq:shannon entropy}) in terms of a relative density
$\tilde{u}=u/u_{*}$. The equal sign of eq.(\ref{eq:H-theorem}) can
only be achieved when the Shannon entropy $N(M^{4},\tau)$ finally
flows to the extreme or ``thermo-equilibrium'' value $N(M_{*}^{4})$.
In the situation, the spacetime manifold $M^{4}$ flows to a final
gradient shrinking Ricci soliton configuration $M_{*}^{4}$ satisfying
\begin{equation}
R_{\mu\nu}-\nabla_{\mu}\nabla_{\nu}\log u=\frac{1}{2\tau}g_{\mu\nu}\label{eq:GSRS}
\end{equation}
The Ricci flow changes both the volume and shape of a spacetime manifold,
but the gradient shrinking Ricci soliton configuration of the spacetime
flows with $\tau$ only change the volume but its shape, so it is
a final ``thermo-equilibrium'' state up to a volume rescaling.

The Ricci flow of the metric conserves the topology of the spacetime
manifold, but blurs the coordinates and hence changes its quadratic
form of distance, so the spacetime under the Ricci flow is non-isotropic.
On the other hand, the quantum coordinate transformation eq.(\ref{eq:integral transformation})
is irreversible and non-unitary. As a consequence, a general quantum
coordinate transformation $X\rightarrow\hat{X}_{\tau}$ suffers from
the diffeomorphism anomaly \citep{Luo:2021zpi}
\begin{equation}
Z(M^{4})\rightarrow Z(\hat{M}^{4})=e^{\lambda N(\hat{M}^{4})}Z(M^{4})\label{eq:Z->Z}
\end{equation}
where $Z(M^{4})=\int\mathscr{D}X\exp\left(-S_{X}\right)$ is the partition
function of the NLSM, $N(M^{4})$ is the Shannon entropy eq.(\ref{eq:shannon entropy}).
The NLSM action $S_{X}$ eq.(\ref{eq:NLSM}) is invariant under the
general coordinate transformation (diffeomorphism), but the path integral
measure $\mathscr{D}X$ changes, which leads to the diffeomorphism
anomaly. A classical diffeomorphism does not change the curvature
of the coordinate, but when the anomaly is taken into account, spacetime
curvature and gravity emerge, which can be seen later. 

Without losing generality, compared with the coordinate-transformed
$\hat{M}^{4}$, let the to-be-transformed $M^{4}$ be the spacetime
with classical coordinates without fluctuation (e.g. laboratory frame).
In this situation, the NLSM action takes a simple form $S_{X}(M^{4})=\frac{1}{2}\lambda\int_{M^{4}}d^{d}xg_{\mu\nu}g^{\mu\nu}=\frac{D}{2}$,
$D=g_{\mu\nu}g^{\mu\nu}\equiv4$ is the dimension of the spacetime,
then we have $Z(M^{4})=e^{-D/2}=e^{-2}$. By using eq.(\ref{eq:Z->Z}),
a partition function normalized by an anomalous counterpart in terms
of the extreme Shannon entropy $N(M_{*}^{4})$ can be given by 
\begin{equation}
\tilde{Z}(\hat{M}^{4},\tau)=\frac{e^{\lambda N(\hat{M}^{4},\tau)-\nu(\hat{M}_{UV}^{4})-2}}{e^{\lambda N(M_{*}^{4})}}=e^{\lambda\tilde{N}(\hat{M}^{4},\tau)-\nu(\hat{M}_{UV}^{4})-2}
\end{equation}
where 
\begin{equation}
\nu(\hat{M}_{UV}^{4})=\lim_{\tau\rightarrow\infty}\lambda\tilde{N}(\hat{M}^{4},\tau)=\lambda\tilde{N}(\hat{M}_{UV}^{4})\label{eq:niu}
\end{equation}
is a constant counter term used to cancel the relative Shannon entropy
$\tilde{N}(\hat{M}^{4},\tau)$ at the laboratory frame up to UV scale
i.e. $\tau\rightarrow\infty$, according to the strong QEP.

A small $\tau$ expansion of the effective action of the partition
function gives
\begin{equation}
-\log\tilde{Z}(\hat{M}^{4})=S_{X(eff)}(\hat{M}^{4})=\lambda\int_{\hat{M}^{4}}dV(X)\,u_{0}\left[2-R(0)\tau+\nu+O(R^{2}\tau^{2})\right],\quad(\tau\rightarrow0)\label{eq:gravity action}
\end{equation}
Noting the Ricci flow of the scalar curvature $R(\tau)=\frac{R(0)}{1+\frac{1}{2}R(0)\tau}$
and the coupling constant of NLSM 
\begin{equation}
\lambda=\frac{3H_{0}^{2}}{8\pi G}=\frac{R(0)}{32\pi G}\label{eq:critial density}
\end{equation}
taking the value of the critical density $\rho_{c}=\frac{3H_{0}^{2}}{8\pi G}$
of the universe, where $H_{0}$ is the Hubble's constant, $G$ the
Newtonian constant and the scalar curvature is $R(0)=D(D-1)H_{0}^{2}=12H_{0}^{2}$.
The effective action is nothing but the Einstein-Hilbert action at
IR ($\tau\rightarrow0$) plus a cosmological constant 
\begin{equation}
S_{X(eff)}(M^{4})=\int_{M^{4}}dV(X)\,\left[\frac{R(\tau)}{16\pi G}+\lambda\nu+O(R^{2}\tau^{2})\right],\quad(\tau\rightarrow0)
\end{equation}
where for a 4-ball $B^{4}$ spacetime $\lambda\nu(B^{4})\approx-0.8\rho_{c}=-\Omega_{\Lambda}\rho_{c}=\frac{-\Lambda}{8\pi G}$
plays the role of a cosmological constant $\Lambda$ being of order
of the critical density $\rho_{c}$ of the universe.

Note in eq.(\ref{eq:gravity action}) that the second-order moment
quantum fluctuation of the material quantum reference system replaces
the $2\lambda$ term in eq.(\ref{eq:1st action}) by $2\lambda-\lambda R(0)\tau+\lambda\nu=\frac{R(\tau)}{16\pi G}+\lambda\nu$.
So in this case, the first-order moment approximation action eq.(\ref{eq:1st action})
is modified to the second-order moment approximation 
\begin{equation}
S[\phi,X]\overset{2nd}{\approx}S[\phi(X)]=\int dV(X)\left[\frac{1}{2}g^{\mu\nu}\frac{\delta\phi}{\delta X^{\mu}}\frac{\delta\phi}{\delta X^{\nu}}-V(\phi)+\frac{R(\tau)}{16\pi G}+\lambda\nu\right],\quad(\tau\rightarrow0)
\end{equation}
which is nothing but a standard action of scalar+gravity plus a cosmological
constant $\lambda\nu$. Einstein's gravity theory emerges from considering
the second-order moment quantum correction (Ricci flow correction)
of the material quantum reference system. So the idea of quantum reference
system may shed new light to the problem of quantum gravity as we
would hope in our previous works \citep{Luo2014The,Luo2015Dark,Luo:2015pca,Luo:2019iby,Luo:2021zpi,Luo:2022goc,Luo:2022statistics,Luo:2022ywl,Luo:2023eqf,2023AnPhy.45869452L,2024chinarxiv}.

\section*{Appendix B: A Modified Distance-Redshift Relation from the Universal
Second-Order Moment Quantum Fluctuation and the Cosmic Acceleration
Expansion}

Through the observations of Type-Ia supernovae at redshift ($z\sim0.5-1$),
the distance-redshift relation is one of the observational approaches
to discover the cosmic acceleration expansion and the evidence of
the existence of ``dark energy'' or a positive cosmological constant.
Because of the fine-tuning nature of the cosmological constant at
the quantum level, it becomes a big mysterious to the fundamental
physics. In the appendix A, we have shown that the cosmological constant
arises as a counter term to the diffeomorphism anomaly coming from
the second-order moment quantum fluctuation of the spacetime (quantum
reference frame). At the phenomenological level, the effect can also
be viewed as a modification of the distance-redshift relation at the
quadratic order from the second-order moment quantum fluctuation of
spacetime \citep{Luo2015Dark,Luo:2021zpi,Luo:2023eqf}, while the
zero-point vacuum energy as a dynamic part plays no role in it. 

In a Friedmann-Robertson-Walker metric at late epoch, when space and
time are on an equal footing,
\begin{equation}
ds^{2}=a^{2}\left(-dt^{2}+dx^{2}+dy^{2}+dz^{2}\right)\label{eq:FRW}
\end{equation}
As a consequence, the Ricci flow of the metric eq.(\ref{eq:Ricci flow})
is simplified to the Ricci flow of the scale factor $a^{2}$, $\frac{da^{2}}{dt}=-2(D-1)$,
where the Ricci curvature is $R_{\mu\nu}=\frac{(D-1)}{a^{2}}g_{\mu\nu}$,
and $D\equiv4$ is the dimension of the spacetime. Clearly, the flow
equation shrinks the scale factor to a singular point at finite flow
parameter, so we need the counter term $\nu$, eq.(\ref{eq:niu}),
which keeps its volume at the laboratory frame up to UV scale and
prevents the metric shrinking to a singular point.

Considering the second-order moment fluctuation eq.(\ref{eq:2nd moment}),
it introduces a second-order metric correction $\langle\delta g_{(2)}^{\mu\nu}\rangle$
to the first-order metric $g_{(1)}^{\mu\nu}$, i.e. $\langle g^{\mu\nu}\rangle=g_{(1)}^{\mu\nu}+\langle\delta g_{(2)}^{\mu\nu}\rangle$,
and also note that $\langle\delta g_{(2)}^{\mu\nu}\rangle=-\langle\delta g_{\mu\nu}^{(2)}\rangle=-\langle\delta a^{2}\rangle$,
we find that the change of the scale factor $\langle\Delta a^{2}\rangle$
during a finite redshift interval $\Delta z$ and a finite flow change
$\Delta\tau$ is given by $\langle\Delta a^{2}\rangle=\langle\Delta a\rangle^{2}-\langle\delta a^{2}\rangle$.
$\langle\Delta a\rangle$ is the change of its mean value leading
to the redshift of the cosmic spectral lines, $\langle\delta a^{2}\rangle$
is given by the Ricci flow or the second-order moment quantum fluctuation
around the mean value, leading to the broadening of the spectral lines.
At IR limit $\tau\rightarrow0$,
\begin{equation}
\lim_{\tau\rightarrow0}\frac{\langle\delta a^{2}\rangle}{\langle\Delta a\rangle^{2}}=1-e^{\frac{2}{D}\nu}\approx0.33,\label{eq:a ratio}
\end{equation}
where $\nu\approx-0.8$ is the counter term of the Ricci flow of the
metric eq.(\ref{eq:FRW}) canceling its volume shrinking, and $e^{\nu}=\lim_{\tau\rightarrow\infty}\frac{\textrm{Vol}(B^{4})}{\textrm{Vol}(R^{4})}$
measures a limit volume comparison between the curved metric eq.(\ref{eq:FRW})
and a flat standard one. The ratio eq.(\ref{eq:a ratio}) between
the second-order moment fluctuation (variance) and the squared first-order
moment (mean square) of the scale factor can be translated to the
redshift version,
\begin{equation}
\lim_{\tau\rightarrow0}\frac{\langle\delta z^{2}\rangle}{\langle z\rangle^{2}}\approx2\left(1-e^{\frac{2}{D}\nu}\right)\approx0.67.\label{eq:z ratio}
\end{equation}
When one considers the Taylor expansion of the luminosity distance
$d_{L}$ by the powers of the redshift $z$, the distance-redshift
relation is given by
\begin{equation}
d_{L}=\frac{1}{H_{0}}\left(z+\frac{1}{2}z^{2}+......\right),\quad(z\rightarrow0).
\end{equation}
The second-order moment quantum fluctuation or broadening of the spectral
lines and the redshift $\langle\delta z^{2}\rangle$ does not modify
its first-order term, but the quadratic-order term by $\langle z^{2}\rangle=\langle z\rangle^{2}+\langle\delta z^{2}\rangle$,
so
\begin{equation}
d_{L}=\frac{1}{H_{0}}\left[\langle z\rangle+\frac{1}{2}\left(1-q_{0}\right)\langle z\rangle^{2}+......\right],\quad(z\rightarrow0)
\end{equation}
where 
\begin{equation}
q_{0}=-\frac{\langle\delta z^{2}\rangle}{\langle z\rangle^{2}}\approx-0.67
\end{equation}
is a calculated deceleration parameter consistent with observational
fittings. 

The modification at the quadratic order can be observed at sufficient
redshift ($z\sim0.5-1$) through the Type-Ia supernovae. The fitting
of the (negative) deceleration parameter actually is a main evidence
of the comic acceleration expansion and the evidence of the positive
cosmological constant. From the viewpoint of the paper, it is very
probably originated from the universal second-order moment quantum
fluctuation/fuzziness of the spacetime (scale factor $\langle\delta a^{2}\rangle$).
And this conjecture is needed to be carefully tested by a direct measurement
of the eq.(\ref{eq:z ratio}), besides the quadratic order correction
of the distance-redshift relation. It is also worth stressing that,
as is shown in this Appendix, the spacetime fuzziness is equivalent
to the (cosmic) acceleration and gravitation (here is repulsion) as
is claimed by the QEP.\bibliographystyle{plain}

\begin{thebibliography}{10}

\bibitem{2020Quantum}
H.~Albers et~al.
\newblock {Quantum test of the Universality of Free Fall using rubidium and
  potassium}.
\newblock {\em Eur. Phys. J. D}, 74(7):145, 2020.

\bibitem{Ali:2006ub}
Md.~Manirul Ali, A.~S. Majumdar, Dipankar Home, and Alok~Kumar Pan.
\newblock {On the quantum analogue of Galileo's leaning tower experiment}.
\newblock {\em Class. Quant. Grav.}, 23:6493--6502, 2006.

\bibitem{1996PhRvD..54.7097A}
C.~{Alvarez} and R.~B. {Mann}.
\newblock {Equivalence principle and anomalous magnetic moment experiments}.
\newblock {\em \prd}, 54(12):7097--7107, December 1996.

\bibitem{1996MPLA...11.1757A}
C.~{Alvarez} and R.~B. {Mann}.
\newblock {Test of the Equivalence Principle Using Atomic Vacuum Energy
  Shifts}.
\newblock {\em Modern Physics Letters A}, 11(21):1757--1763, January 1996.

\bibitem{1996PhRvD..54.5954A}
C.~{Alvarez} and R.~B. {Mann}.
\newblock {Testing the equivalence principle by Lamb shift energies}.
\newblock {\em \prd}, 54(10):5954--5974, November 1996.

\bibitem{Anastopoulos_2018}
C~Anastopoulos and B~L Hu.
\newblock Equivalence principle for quantum systems: dephasing and phase shift
  of free-falling particles.
\newblock {\em Classical and Quantum Gravity}, 35(3):035011, jan 2018.

\bibitem{ALPHA:2023dec}
E.~K. Anderson et~al.
\newblock {Observation of the effect of gravity on the motion of antimatter}.
\newblock {\em Nature}, 621(7980):716--722, 2023.

\bibitem{2022Testing}
Brynle Barrett et~al.
\newblock {Testing the universality of free fall using correlated
  39K\textendash{}87Rb atom interferometers}.
\newblock {\em AVS Quantum Sci.}, 4(1):014401, 2022.

\bibitem{1979AmJPh..47..264B}
M.~V. {Berry} and N.~L. {Balazs}.
\newblock {Nonspreading wave packets}.
\newblock {\em American Journal of Physics}, 47(3):264--267, March 1979.

\bibitem{2022Free}
A.~Colcelli, G.~Mussardo, G.~Sierra, and A.~Trombettoni.
\newblock Free fall of a quantum many-body system.
\newblock {\em American Journal of Physics}, 90(11):833--840, 2022.

\bibitem{1975Observation}
R.~Colella, A.~W. Overhauser, and S.~A. Werner.
\newblock Observation of gravitationally induced quantum interference.
\newblock {\em Physical Review Letters}, 34(23):1472--1474, 1975.

\bibitem{2015Spinning}
E.~Corinaldesi and A.~Papapetrou.
\newblock Spinning test-particles in general relativity. ii.
\newblock {\em Physics Today}, 98(3):317--325, 2015.

\bibitem{2004CQGra..21.2761D}
P.~C.~W. {Davies}.
\newblock {Quantum mechanics and the equivalence principle}.
\newblock {\em Classical and Quantum Gravity}, 21(11):2761--2772, June 2004.

\bibitem{Eliezer1977The}
Eliezer and J.~C.
\newblock The equivalence principle and quantum mechanics.
\newblock {\em American Journal of Physics}, 45(12):1218--1221, 1977.

\bibitem{2022On}
Viacheslav~A. Emelyanov.
\newblock {On free fall of quantum matter}.
\newblock {\em Eur. Phys. J. C}, 82(4):318, 2022.

\bibitem{2020arXiv201213754G}
Flaminia {Giacomini} and {\v{C}}aslav {Brukner}.
\newblock {Einstein's Equivalence principle for superpositions of gravitational
  fields and quantum reference frames}.
\newblock {\em arXiv e-prints}, page arXiv:2012.13754, December 2020.

\bibitem{2022Quantum}
Flaminia Giacomini and \v{C}aslav Brukner.
\newblock {Quantum superposition of spacetimes obeys Einstein's equivalence
  principle}.
\newblock {\em AVS Quantum Sci.}, 4(1):015601, 2022.

\bibitem{Giacomini:2017zju}
Flaminia Giacomini, Esteban Castro-Ruiz, and \v{C}aslav Brukner.
\newblock {Quantum mechanics and the covariance of physical laws in quantum
  reference frames}.
\newblock {\em Nature Commun.}, 10(1):494, 2019.

\bibitem{Greenberger1970Theory}
Greenberger and M.~Daniel.
\newblock Theory of particles with variable mass. ii. some physical
  consequences.
\newblock {\em Journal of Mathematical Physics}, 11(8):2341--2347, 1970.

\bibitem{Greenberger1998Some}
Greenberger and M.~Daniel.
\newblock Some remarks on the extended galilean transformation.
\newblock {\em American Journal of Physics}, 47(1):35--38, 1998.

\bibitem{1983The}
Daniel~M. Greenberger.
\newblock The neutron interferometer as a device for illustrating the strange
  behavior of quantum systems.
\newblock {\em Review of Modern Physics}, 55(4):875--905, 1983.

\bibitem{Herdegen:2001tt}
Andrzej Herdegen and Jaroslaw Wawrzycki.
\newblock {Is Einstein's equivalence principle valid for a quantum particle?}
\newblock {\em Phys. Rev. D}, 66:044007, 2002.

\bibitem{2006Quantum}
Stella Huerfano, Sarira Sahu, and M.~Socolovsky.
\newblock Quantum mechanics and the weak equivalence principle.
\newblock {\em Int.j.pure Appl.math}, (2):153--166, 2006.

\bibitem{2012Weak}
Clovis {Jacinto de Matos}.
\newblock {Weak Equivalence Principle and Propagation of the Wave Function in
  Quantum Mechanics}.
\newblock {\em arXiv e-prints}, page arXiv:1006.2657, June 2010.

\bibitem{1966Can}
M.~Kac.
\newblock Can one hear the shape of drum?
\newblock {\em American Mathematics Monthly}, 73, 1966.

\bibitem{Kucha1980Gravitation}
Kucha and Karel.
\newblock Gravitation, geometry, and nonrelativistic quantum theory.
\newblock {\em Physical Review D}, 22(6):1285--1299, 1980.

\bibitem{Luo2014The}
M.~J. Luo.
\newblock The cosmological constant problem and re-interpretation of time.
\newblock {\em Nuclear Physics}, 884(1):344--356, 2014.

\bibitem{Luo2015Dark}
M.~J. Luo.
\newblock Dark energy from quantum uncertainty of distant clock.
\newblock {\em Journal of High Energy Physics}, 06(063):1--11, 2015.

\bibitem{Luo:2015pca}
M.~J. Luo.
\newblock The cosmological constant problem and quantum spacetime reference
  frame.
\newblock {\em Int. J. Mod. Phys.}, D27(08):1850081, 2018.

\bibitem{Luo:2019iby}
M.~J. Luo.
\newblock {Ricci Flow Approach to The Cosmological Constant Problem}.
\newblock {\em Found. Phys.}, 51(1):2, 2021.

\bibitem{Luo:2021zpi}
M.~J. Luo.
\newblock {Trace anomaly, Perelman\textquoteright{}s functionals and the
  cosmological constant}.
\newblock {\em Class. Quant. Grav.}, 38(15):155018, 2021.

\bibitem{Luo:2022goc}
M.~J. Luo.
\newblock {Local conformal instability and local non-collapsing in the Ricci
  flow of quantum spacetime}.
\newblock {\em Annals Phys.}, 441:168861, 2022.

\bibitem{Luo:2022statistics}
M.~J. Luo.
\newblock {A Statistical Fields Theory underlying the Thermodynamics of Ricci
  Flow and Gravity}.
\newblock {\em Int. J. Mod. Phys. D}, 32(5):2350022, 2 2023.

\bibitem{Luo:2022ywl}
M.~J. Luo.
\newblock {Quantum Modified Gravity at Low Energy in the Ricci Flow of Quantum
  Spacetime}.
\newblock {\em Int. J. Theor. Phys.}, 62(4):91, 2023.

\bibitem{2023AnPhy.45869452L}
M.~J. {Luo}.
\newblock {The Ricci flow and the early universe}.
\newblock {\em Annals of Physics}, 458:169452, November 2023.

\bibitem{2024chinarxiv}
M.~J. {Luo}.
\newblock {A Theory of Quantum Reference Frame and its Implications to
  Gravity}.
\newblock {\em ChinaXiv:202404.00156 (in Chinese)}, 2024.

\bibitem{Luo:2023eqf}
M.~J. Luo.
\newblock {Local Short-Time Acceleration induced Spectral Line Broadening and
  Possible Implications in Cosmology}.
\newblock {\em Annals of Physics}, 473:169899, November 2024.

\bibitem{Nauenberg2016Einstein}
Nauenberg and Michael.
\newblock Einstein's equivalence principle in quantum mechanics revisited.
\newblock {\em American Journal of Physics}, 84(11):879--882, 2016.

\bibitem{Okon:2010kn}
Elias Okon and Craig Callender.
\newblock {Does Quantum Mechanics Clash with the Equivalence Principle - and
  Does it Matter?}
\newblock {\em Eur. J. Phil. Sci.}, 1:133--145, 2011.

\bibitem{1974Experimental}
A.~W. Overhauser and R.~Colella.
\newblock Experimental test of gravitationally induced quantum interference.
\newblock {\em Physical Review Letters}, 33(20):1237--1239, 1974.

\bibitem{2017Does}
Luigi Seveso, Valerio Peri, and Matteo G.~A. Paris.
\newblock {Does universality of free-fall apply to the motion of quantum
  probes?}
\newblock {\em J. Phys. Conf. Ser.}, 880(1):012067, 2017.

\bibitem{Vandegrift2000Accelerating}
Vandegrift and G.
\newblock Accelerating wave packet solution to schrodinger's equation.
\newblock {\em American Journal of Physics}, 68(6):576--577, 2000.

\bibitem{1979GReGr..10..181V}
R.~F.~C. {Vessot} and M.~W. {Levine}.
\newblock {A test of the equivalence principle using a space-borne clock}.
\newblock {\em General Relativity and Gravitation}, 10(3):181--204, February
  1979.

\bibitem{1997Testing}
Lorenza Viola and Roberto Onofrio.
\newblock Testing the equivalence principle through freely falling quantum
  objects.
\newblock {\em Physical Review D}, 55(2):455--455, 1997.

\bibitem{Wadati2000The}
Wadati and Miki.
\newblock The free fall of quantum particles.
\newblock {\em J Phys Soc Jpn}, 68(8):2543--2546, 2000.

\bibitem{Wilczynska:2020rxx}
Michael~R. Wilczynska et~al.
\newblock {Four direct measurements of the fine-structure constant 13 billion
  years ago}.
\newblock {\em Sci. Adv.}, 6(17):eaay9672, 2020.

\bibitem{2017Quantum}
Magdalena Zych and \v{C}aslav Brukner.
\newblock {Quantum formulation of the Einstein Equivalence Principle}.
\newblock {\em Nature Phys.}, 14(10):1027--1031, 2018.

\end{thebibliography}

\end{document}